\documentclass[11pt]{article}%
\usepackage{amsfonts}
\usepackage{amsmath}
\usepackage{amssymb}
\usepackage{graphicx}
\usepackage{hyperref}
\usepackage{textcomp}
\usepackage{amstext}%
\providecommand{\U}[1]{\protect\rule{.1in}{.1in}}
\begin{document}

\title{The Doctoral Students of Richard Feynman}
\author{{\Large T. S. Van Kortryk\medskip}\\120 Payne Street, Paris MO65275\\{\small vankortryk@gmail.com}}
\date{}
\maketitle

\begin{abstract}
I document 35 students who graduated to receive PhDs under Feynman's
supervision. \ I provide links to their doctoral dissertations.

\end{abstract}

\vfill

\section*{Introduction\smallskip}

\begin{quote}
An ordinary genius is an ordinary fellow ... There is no mystery as to how his
mind works. ... It is different with the magicians ... Even after we
understand what they have done, the process by which they have done it is
completely dark. \ \emph{They seldom, if ever, have students} ...\ Richard
Feynman is a magician of the highest caliber.\ 

\hfill--- Mark Kac \cite{Kac}\smallskip
\end{quote}

Since this is the centennial year of
\href{https://en.wikipedia.org/wiki/Richard_Feynman}{Richard Feynman}'s birth,
I attempt here to dispel a minor myth about\ him. \ The myth is embodied in
the words of \href{https://en.wikipedia.org/wiki/Mark_Kac}{Mark Kac} that I
have emphasized above, and in the following statement attributed to one of
Feynman's students, Philip Platzman\ \cite{Mehra}: \textquotedblleft The
reason why Feynman did not have many students was because he was very
difficult with them, because he didn't really worry about students. ...\ He
had a few students, but not many.\textquotedblright\ \ 

Thus a prevailing belief in the scientific community seems to be \cite{TLC}
\emph{Feynman had very few doctoral students who completed theses under his
supervision}. \ It may be surprising to most people to learn this is
\emph{not} true.

The number of Feynman's doctoral students is actually $\mathbf{35~}%
\boldsymbol{\pm~3}$, with the uncertainty intended to take into account some
unavailable documents as well as possible subjectivity on my part
\cite{Subjectivity Footnote}. \ The lineup of students who completed their PhD
research under Feynman's discerning gaze began in 1951 with Michel Baranger,
\href{https://en.wikipedia.org/wiki/Laurie_Brown_(physicist)}{Laurie Brown},
and \href{https://en.wikipedia.org/wiki/Giovanni_Rossi_Lomanitz}{Giovanni
Lomanitz} at Cornell, and continued until at least 1977 with Ted Barnes and
\href{https://en.wikipedia.org/wiki/Thomas_Curtright}{Thomas Curtright} at Caltech.

While $\boldsymbol{35}$ is not an extremely large number of doctoral students
to have mentored during a lifetime as an academic (e.g. compared to the
$\boldsymbol{70+}$ dissertations supervised by
\href{https://en.wikipedia.org/wiki/Julian_Schwinger}{Julian Schwinger} during
five decades), nonetheless, $\boldsymbol{35}$ does amount on average to almost
one PhD for every year of Feynman's time as a professor. \ Moreover, were it
not for illness during the last several years of his career, Feynman might
have supervised several more students.

Among Feynman's doctoral students the most recognized physicist is undoubtedly
\href{https://en.wikipedia.org/wiki/George_Zweig}{George Zweig}, who was also
mentored by \href{https://en.wikipedia.org/wiki/Murray_Gell-Mann}{Murray
Gell-Mann} and \href{https://en.wikipedia.org/wiki/Alvin_V._Tollestrup}{Alvin
Tollestrup}, and who graduated from Caltech in 1964. \ Soon thereafter Zweig
had a major impact on elementary particle physics through his independent
invention of the \textquotedblleft quark model\textquotedblright\ of hadrons.
\ However, in my opinion, the research of Feynman's other students has also
had significant impact and continues to influence several areas of
physics.\newpage

\section*{The Students}

\begin{quote}
There are PhDs and then there are Feynman PhDs. --- Richard
Sherman\ \cite{Mehra}
\end{quote}

From theses and PhD dissertation examination documents wherein it was either
explicitly stated or otherwise clear that Feynman was the advisor$^{\ast}$ or
co-advisor$^{\ast\ast}$, I find the $\boldsymbol{35}$ doctoral students listed
here, the first three at Cornell, the others at Caltech:\bigskip

Michel Baranger$^{\ast\ast}$ (1951)
\href{https://newcatalog.library.cornell.edu/catalog/8545}{Relativistic
Corrections to the Lamb Shift}

Laurie Brown$^{\ast}$ (1951)
\href{https://newcatalog.library.cornell.edu/catalog/12688}{Radiative
corrections to the Klein-Nishina formula}

Giovanni Lomanitz$^{\ast}$ (1951)
\href{https://newcatalog.library.cornell.edu/catalog/33213}{Second order
effects in the electron-electron interaction}

Carl Wilhelm Hellstrom$^{\ast\ast}$ (1951)
\href{https://thesis.library.caltech.edu/10515/}{Production and Annihilation
of Antiprotons}

Howard Murray Robbins$^{\ast\ast}$ (1952)
\href{https://thesis.library.caltech.edu/10568/}{I. Retardation Corrections
... II. Self Energy ...}

Albert Hibbs$^{\ast}$ (1955)
\href{https://thesis.library.caltech.edu/5005/}{The growth of water waves due
to the action of the wind}

William Karzas$^{\ast\ast}$ (1955)
\href{https://thesis.library.caltech.edu/45/}{The effects of atomic electrons
on nuclear radiation}

Koichi Mano$^{\ast}$ (1955) \href{https://thesis.library.caltech.edu/167/}{The
self-energy of the scalar nucleon}

Gerald Speisman$^{\ast}$ (1955)
\href{https://thesis.library.caltech.edu/264/}{The neutron-proton mass
difference}

Truman Woodruff$^{\ast\ast}$ (1955)
\href{https://thesis.library.caltech.edu/365/}{On the orthogonalized plane
wave method for calculating ...}

Michael Cohen$^{\ast}$ (1956)
\href{https://thesis.library.caltech.edu/1007/}{The energy spectrum of the
excitations in liquid helium}

Samuel Berman$^{\ast}$ (1959)
\href{https://thesis.library.caltech.edu/380/}{Radiative corrections to muon
and neutron decay}

Frank Vernon$^{\ast}$ (1959)
\href{https://thesis.library.caltech.edu/737/}{The theory of a general quantum
system interacting ... dissipative system}

Willard Wells$^{\ast}$ (1959)
\href{https://thesis.library.caltech.edu/708/}{Quantum theory of coupled
systems having application to masers}

Henry Hilton$^{\ast\ast}$ (1960)
\href{https://thesis.library.caltech.edu/2687/}{Comparison of the beta-spectra
of boron 12 and nitrogen 12}

Carl Iddings$^{\ast}$ (1960)
\href{https://thesis.library.caltech.edu/2654/}{Nuclear size corrections to
the hyperfine structure of hydrogen}

Philip Platzman$^{\ast\ast}$ (1960)
\href{https://thesis.library.caltech.edu/2738/}{Meson theoretical origins of
the non-static two nucleon potential}

Marvin Chester$^{\ast\ast}$ (1961)
\href{https://thesis.library.caltech.edu/985/}{Some experimental and
theoretical observations on ... EMF}

Elisha Huggins$^{\ast}$ (1962)
\href{https://thesis.library.caltech.edu/6592/}{Quantum mechanics of the
interaction of gravity ...}

Harold Yura$^{\ast}$ (1962)
\href{https://thesis.library.caltech.edu/6594/}{The quantum electrodynamics of
a medium}

Michael Levine$^{\ast\ast}$ (1963)
\href{https://thesis.library.caltech.edu/7363/}{Neutrino processes of
significance in stars}

George Zweig$^{\ast\ast}$ (1964)
\href{https://thesis.library.caltech.edu/4937/}{Two topics in elementary
particle physics ...}

James Bardeen$^{\ast\ast}$ (1965)
\href{https://thesis.library.caltech.edu/1392/}{Stability and dynamics of
spherically symmetric masses ...}

Richard William Griffith$^{\ast\ast}$ (1969)
\href{https://thesis.library.caltech.edu/10108/}{Chiral Symmetry Breaking:
Meson and Nucleon Masses}

Howard Arthur Kabakow$^{\ast\ast}$ (1969)
\href{https://thesis.library.caltech.edu/7605/}{A perturbation procedure for
nonlinear oscillations ...}

Robert Carlitz$^{\ast\ast}$ (1970)
\href{https://thesis.library.caltech.edu/9562/}{Elimination of parity doubled
states from Regge amplitudes} \cite{Carlitz Footnote}

Mark Kislinger$^{\ast\ast}$ (1970)
\href{https://thesis.library.caltech.edu/9600/}{Elimination of parity doublets
in Regge amplitudes}

E. William Colglazier, Jr.$^{\ast\ast}$ (1971)
\href{https://thesis.library.caltech.edu/10820/}{Two Topics in Elementary
Particle Physics}

Finn Ravndal$^{\ast\ast}$ (1971)
\href{https://thesis.library.caltech.edu/1574/}{A relativistic quark model
with harmonic dynamics} \cite{Ravndal Footnote}

Richard Sherman$^{\ast}$ (1971)
\href{https://thesis.library.caltech.edu/9599/}{Surface impedance theory for
superconductors in ... magnetic fields}

Arturo Cisneros$^{\ast\ast}$ (1973)
\href{https://thesis.library.caltech.edu/9601/}{I. Baryon-Antibaryon phase
transition ... \ II. ... the Parton Model}

Steven Kauffmann$^{\ast}$ (1973)
\href{https://thesis.library.caltech.edu/8178/}{Ortho-positronium annihilation
... first order radiative corrections}

Robert Wang$^{\ast\ast}$ (1976)
\href{https://thesis.library.caltech.edu/10583/}{A Study of Some
Two-Dimensional Field Theory Models}

Frank (Ted) Barnes$^{\ast\ast}$ (1977)
\href{https://thesis.library.caltech.edu/3543/}{Quarks, gluons, bags, and
hadrons}

Thomas L. Curtright$^{\ast}$ (1977)
\href{https://thesis.library.caltech.edu/4898/}{Stability and Supersymmetry}%
\bigskip

From theses where Feynman was \emph{not} described as an advisor or
co-advisor, but was a member of the PhD examination committee although not the
committee chairman, and/or was acknowledged in the work for moderate influence
and/or general advice, I find in addition:\bigskip

Fredrik Zachariasen (1956)
\href{https://thesis.library.caltech.edu/2041/}{Photodisintegration of the
deuteron}

Paul Craig (1959) \href{https://thesis.library.caltech.edu/455/}{Observations
of perfect potential flow and critical velocities in superfluid ...}

James Mercereau (1959)
\href{https://thesis.library.caltech.edu/593/}{Diffraction of Thermal Waves in
Liquid Helium II}

Kenneth Wilson (1961) \href{https://thesis.library.caltech.edu/4205/}{An
investigation of the Low ... and the Chew-Mandelstam equations}

John Andelin (1966) \href{https://thesis.library.caltech.edu/2383/}{Superfluid
drag in helium II}

Karvel Thornber (1966) \href{https://thesis.library.caltech.edu/9247/}{I.
Electronic Processes ... II. Polaron Motion ...}

Lorin Vant-Hull (1967)
\href{https://thesis.library.caltech.edu/3705/}{Verification of long range
quantum phase coherence ...}

William Press (1973)
\href{https://thesis.library.caltech.edu/3221/}{Applications of black-hole
perturbation techniques}

Don Page (1976) \href{https://thesis.library.caltech.edu/7179/}{Accretion into
and emission from black holes}

Stephen Wolfram (1980)
\href{https://resolver.caltech.edu/CaltechETD:etd-06142007-142536}{Some topics
in theoretical high-energy physics}\bigskip

\noindent All of these were Caltech students. \ Originally I included Platzman
in this second list. \ But upon looking at other documents I became convinced
that his thesis was effectively co-supervised by Feynman to the extent that he
belonged in the first list. \ (Indeed, as I document later, Platzman's
personal listing in the
\href{http://genealogy.math.ndsu.nodak.edu/index.php}{Mathematics Genealogy
Project} states that Feynman was a co-advisor.) \ Similar remarks apply for
Robert Carlitz \cite{Carlitz Footnote} and Finn Ravndal \cite{Ravndal
Footnote}. \ If so, that would justify my head\ count of $\boldsymbol{35}$
\textquotedblleft Feynman PhDs\textquotedblright\ to be a lower bound. \ 

Finally, I find several less compelling cases where Feynman was only a member
of the dissertation examination committee at Caltech and was not particularly
influential for the research, so far as I can tell. \ I suspect there are many
more such cases that I have not found, since on this point documentation is
quite often incomplete and all committee members are not listed. \ For
example:\bigskip

Lipes, Richard Gwin (1969)
\href{http://resolver.caltech.edu/CaltechTHESIS:02182014-114114913}{I.
Application of multi-Regge theory ... II. High energy model ...}.

Hill, Christopher Thaddeus (1977)
\href{http://resolver.caltech.edu/CaltechETD:etd-11112004-110424}{Higgs
scalars and the nonleptonic weak interactions}.

Dally, William J. (1986)
\href{http://resolver.caltech.edu/CaltechETD:etd-03252008-140428}{A VLSI
architecture for concurrent data structures}. \ (restricted)

Wawrzynek, John (1987)
\href{http://resolver.caltech.edu/CaltechETD:etd-03052008-112515}{VLSI
concurrent computation for music synthesis}. \ (restricted)\bigskip\ \ 

\noindent For the last two cases given above, I cannot access the theses to
see if Feynman was acknowledged for significant influence.

\section*{Sources}

At the time of this writing, wikipedia lists only six students to have
officially received PhDs with Feynman as the advisor, in alphabetical order:
\ James M. Bardeen, Laurie Brown, Thomas Curtright, Al Hibbs, Giovanni Rossi
Lomanitz, and George Zweig. \ However, this list is obviously far from
complete, as documented by the Math Genealogy Project (MGP) and by the Caltech
library archives. \ 

\href{http://genealogy.math.ndsu.nodak.edu/id.php?id=91222}{According to the
MGP}, also at the time of this writing, there were at least thirteen other
doctoral degrees completed under Feynman's supervision in addition to those
listed in wikipedia. \ In particular, Philip Platzman is in the MGP list but
not in wikipedia. I suppose that is because he requested MGP to classify him
as a doctoral student of Feynman.
\ \href{http://genealogy.math.ndsu.nodak.edu/id.php?id=136252}{Platzman's
personal listing in MGP} supports my supposition. \ By way of comparison, and
as a measure of the completeness of their database,
\href{http://genealogy.math.ndsu.nodak.edu/id.php?id=15198}{Schwinger} has
only 21 of his students listed by the MGP. \ 

In any case, the mother lode of information about Feynman's doctoral students
can be found at \href{http://thesis.library.caltech.edu/}{the Caltech
library}.
\ \href{http://thesis.library.caltech.edu/view/advisor/Feynman-R-P.html}{A
direct search of their online database} produces a list of 31 PhD theses where
Feynman is described as the advisor or co-advisor, at the time of this writing
\cite{Latest}. \ By way of comparison, a direct search for
\href{http://thesis.library.caltech.edu/view/advisor/Gell-Mann-M.html}{Gell-Mann
as advisor} turns up 18 theses in the Caltech library database. \ Among these,
Hilton and Levine are shown to be co-advised by Feynman and Gell-Mann.
\ Remarkably, Zweig is \emph{not} listed as a Gell-Mann advisee.

Beyond these publicly accessible sources, the largest amount of documentation
that is available to me concerns Thomas Curtright, who has provided this
succinctly amusing excerpt from his thesis examination committee papers
\cite{TLC}:%

{\includegraphics[
height=1.2047in,
width=4.459in
]%
{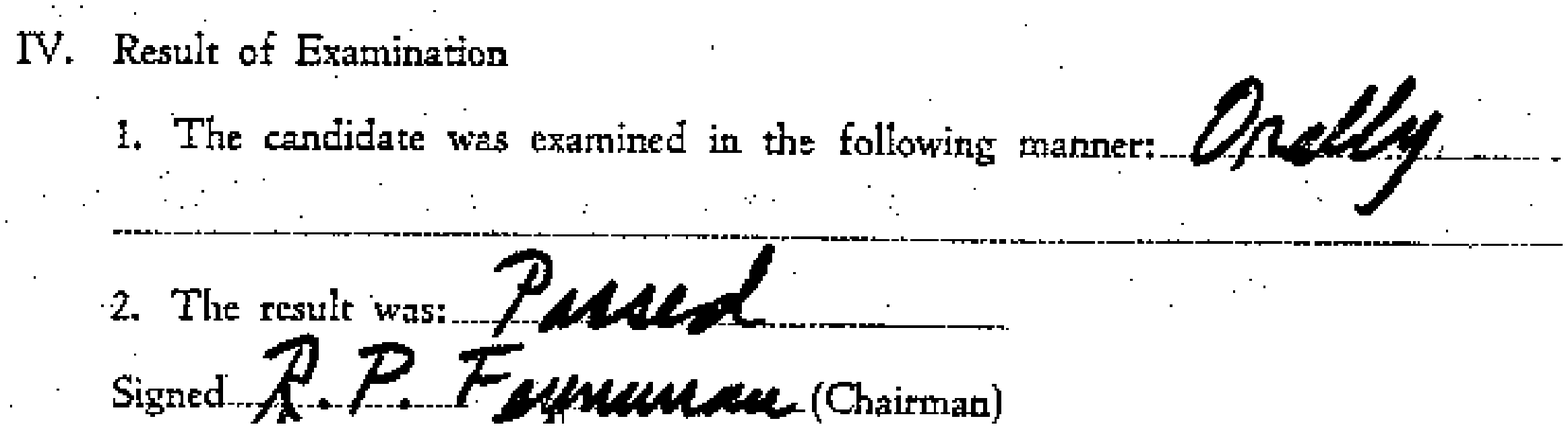}%
}

\section*{Summary}

From looking at many theses and papers by Caltech students, my overall
impression is simply this: \ Feynman played a \emph{major role} through his
mentoring and supervision of doctoral students. \ He exerted tremendous
influence on graduate student research conducted at Caltech during his four
decades there --- perhaps even more than his widely perceived influence on
Caltech undergraduate studies. \ I conclude that it is \emph{not} true Richard
Feynman \textquotedblleft had a few students, but not many.\textquotedblright%
\bigskip

\noindent\textbf{Acknowledgements: \ }I thank Professor Curtright for
suggesting that there could very well be a widespread misunderstanding about
the extent of Feynman's mentoring of doctoral students. \ Finally, I thank
\href{https://en.wikipedia.org/wiki/Cosmas_Zachos}{Cosmas Zachos} for his
comments on various drafts of this manuscript.

\end{document}